\documentclass[a4paper,twcolumn]{IEEEtran}
\usepackage{color}
\usepackage{graphicx}
\usepackage{amssymb}
\usepackage{multicol}
\usepackage{amsmath}
\usepackage{amsthm}
\usepackage{psfrag}
 \usepackage{eso-pic}
\usepackage{mathtools}

\begin{document}

\AddToShipoutPictureFG*{
                \AtPageUpperLeft{\put(0,-10){\makebox[\paperwidth][l]{2016 8th International                 Symposium on Telecommunications (IST'2016)}}}
                                                           }

\title{On the Uplink Spectral Efficiency of Full-Duplex Cooperative OFDMA Systems}
\author{Jafar Banar and  S. Mohammad Razavizadeh \\ School of Electrical Engineering \\ Iran University of Science and Technology (IUST)  \\ J\_ banar@Elec.iust.ac.ir, smrazavi@iust.ac.ir }
\maketitle

\begin{abstract}
In this paper, we develop a resource allocation algorithm for uplink of in-band full-duplex (FD) cellular networks. The FD cellular network is assumed to be based on orthogonal frequency division multiple access (OFDMA) and consists of a base station communicating with multiple users. Some of the users in the network act as relay for other users and help them to transmit their data to the base station. These relays are FD and work based on amplify and forward (AF) protocol. By appropriate selection of the relays and optimized allocation of subcarriers and powers to all users, we try to maximize the total sum rate of the network. During this optimization, we also impose some constraints on the users' quality of service (QoS) and power. We propose a new algorithm to select the best relays based on the users' maximum data rate and also use Linear Assignment Problem Jonker-Volgenant (LAPJV) algorithm for subcarrier assignment. It is proved that the resulting optimization problem can be converted to a convex problem, and hence it can be solved by standard numerical methods. The simulation results demonstrate the effect of the proposed scheme on the sum rate and coverage of the network.
\end{abstract}


\section{Introduction}\label{sec:Introduction}
\IEEEPARstart{T}{he} increasing demand for high-speed wireless services in recent years has led to advent of new technologies to improve data rate and coverage area. Among these technologies, in-band full-duplex (FD) systems and cooperative communications based on different relaying methods have attracted lots of interests [1], [2]. In FD systems, wireless terminals can transmit and receive signals at the same time and same frequency band with negligible self-interference (SI) thanks to appropriate SI mitigation methods. Hence, spectral efficiency (SE) achieved in FD systems may become as twice as in half-duplex (HF) ones [1]. Moreover, employing various relaying methods including amplify and forward (AF) and selective relaying (SR) have been studied extensively in literature as an powerful tool to increase the SE in wireless networks. 

A review on previous papers in this issue shows that there are three methods to communicate with users in FD cellular networks. The first method uses direct link between the base station (BS) and users [8]- [10]. In these papers, both uplink and downlink channel are studied and the self-interference at the BS is also taken into account. The second method is to install fixed relays within the BS coverage area. Most of recent works, e.g., [3] and [6] consider this assumption. In those papers, it is considered that there are some FD relays in the network that can operate in AF or DF mode. In addition, these papers only consider downlink transmission. Finally, the third method is to exploit the users as the relays [5], [11]. These papers are based on orthogonal frequency division multiple access (OFDMA) and again only downlink channel is studied. There are also several papers that employ FD technique in cooperative communication [3]- [7]. 
However, most of these papers are in single user scenarios with one source and one destination and some relays between them [4], [7].  

In this paper, we study the problem of sum-rate maximization in the uplink of a full-duplex cooperative OFDMA network. 
Based on the distance to the BS, the users in the network are divided into two groups. The users, which have shorter distance to the BS act as relays for the users in far distance to the BS. Then, we try to optimally allocate powers to all users, and then select the best relay for each users of the far distance. Afterward, the subcarriers are allocated to the users. For subcarrier allocation, we employ Linear Assignment Problem Jonker-Volgenant (LAPJV) algorithm. 

Then, an optimization problem is formulated with sum-rate as the objective function and power and quality of service (in terms of data rate) of each user as its constraints. 
It is shown that this optimization problem can be converted to a convex problem and solved by standard numerical methods.
The simulation results demonstrate the effect of the proposed scheme on the sum rate and coverage of the network. Specifically, for a given total number of users, by adding more users to the relay group, the performance of the network will be improved. 

The remainder of this paper is organized as follows. The system model is introduced in section II. Problem formulation for sum-rate maximization is presented in section III. The simulation results are demonstrated in section IV. Finally, section V concludes the paper.

\section{System Model}\label{sec:SYSTEM MODEL}
\begin{figure}[!t]
\centering
\includegraphics[width=3in]{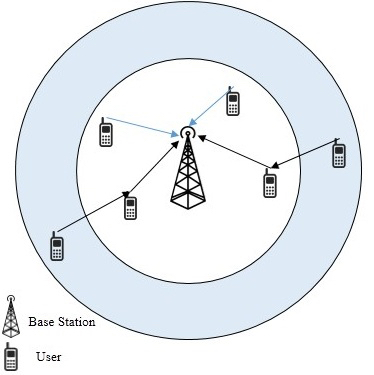}
\caption{System model}
\label{f1}
\end{figure}

As shown in Fig. $\ref{f1}$, we consider the uplink of a cellular network which consists of a BS and two group of users. The number of users of the first group (i.e., the farther users) is $K1$ and for the second group (i.e., the closer users) is $K2$. Total bandwidth of this OFDMA cellular network is equal to $NW$ where $N$ is the total number of subcarriers and $W$ is the bandwidth of each subcarrier, which is assumed to be equal for all subcarriers. All users as well as the BS, operate in FD modes. The users of the first group transmit their data to the users of the second group to relay them to the BS. In fact, first group users are not in the BS coverage area due to their great distances to the BS. Therefore, they need help from some users to communicate with the BS. Then, the users of the second group that are in the coverage and service area of the BS, act as the relay for the users of the first group. These relays use amplify and forward (AF) relaying strategy, in which a relay amplifies and forwards its received signal to the BS. 

To achieve full diversity in selective OFDMA relaying system, we choose the relay with the highest received SINR of the first group users at the BS receiver [12]. That is, in each subcarrier, the best relay is selected based on the maximum signal to interference plus noise (SINR) ratio at the BS. In general, two time slots are needed to transmit data from the users of the first group to the BS. In addition, any user of the second group that is not selected as the relay can send its data to the BS in both time slots. Therefore, the users can be in either cooperative or non-cooperative modes.
We also assumed that all users have a single antenna.

 The channel coefficients between the $k$th user in the first group to the $m$th user in the second group (i.e., relays) in the $i$th subcarrier and from the $m$th user in the second group to the BS in the $j$th subcarrier are denoted by $h_{k,m}^i$ and $g_{m,B}^j$, respectively. In general, in this paper, $i$ and $j$ are the subcarrier superscripts in first and second time slots, respectively.
The channel coefficient matrix from the BS transmit antennas to the BS receive antennas which determines the self-interference is denoted by $H_{SI}$. 

In this section, we formulate the subcarrier and power allocation optimization problem with data rate maximization objective and QoS constraint. As previously mentioned, there are two modes of communication. In cooperative mode, the received signal at the relay in the first time slot is\\

\begin{equation}
\label{e1}
Y_{k,m}^{i,(C,1)} =  \sqrt{|h_{k,m}^i|^2 P_{k,m}^{i,(C,1)}} X_{k,m}^i + Z_{k}^i . \\
\end{equation}

Then, in the second time slot, the received signal at the BS is\\
\begin{equation}
\begin{split}
\label{e2}
Y_{m,B}^{j,(C,2)} &= \sqrt{P_{m,B}^{j,(C,2)} |g_{m,B}^j|^2} G Y_{k,m}^{i,(C,1)} \\
&\quad + \sqrt{P_{B,m}^{j,(C,2)} |H_{SI}^j|^2} X_{B,m}^j + Z_{m}^j\\
&= \sqrt{P_{m,B}^{j,(C,2)} P_{k,m}^{i,(C,1)} |g_{m,B}^j|^2 |h_{k,m}^i|^2} G X_{k,m}^i \\
&\quad+ \sqrt{P_{B,m}^{j,(C,2)} |H_{SI}^j|^2} X_{B,m}^j \\ 
&\quad + \sqrt{P_{m,B}^{j,(C,2)} |g_{m,B}^j|^2} G Z_{k}^i+ Z_{m}^j ,\\ 
\end{split}
\end{equation}

where, $G$ represents the received signal amplification factor at the AF relays and it is equal to\\
\begin{equation}
\begin{aligned}
\label{e3}
G =\frac{1}{ \sqrt{P_{k,m}^{i,(C,1)} |h_{k,m}^i|^2+1}} ,\\
\end{aligned}
\end{equation}

where, $P_{B,m}$ represents transmit power from the BS transmit antennas to its received antennas in the downlink.\\
In non-cooperative mode, the received signal at the BS in the first and second time slots are, respectively, represented as\\

\begin{equation}
\begin{split}
\label{e4}
Y_{m,B}^{i,(NC,1)} &= \sqrt{P_{m,B}^{i,(NC,1)} |g_{m,B}^i|^2} X_{m,B}^i \\
&\quad+ \sqrt{P_{B,m}^{i,(NC,1)} |H_{SI}^i|^2} X_{B,m}^i + Z_{m}^i ,\\
\end{split}
\end{equation}

\begin{equation}
\begin{split}
\label{e5}
Y_{m,B}^{j,(NC,2)} &= \sqrt{P_{m,B}^{j,(NC,2)} |g_{m,B}^j|^2} X_{m,B}^j \\
&\quad+ \sqrt{P_{B,m}^{j,(NC,2)} |H_{SI}^j|^2} X_{B,m}^j  + Z_{m}^j .\\
\end{split}
\end{equation}

The SINR for cooperative communication is derived as ($\ref{e6}$) at the top of the next page. In the denominator of ($\ref{e6}$), we can ignore the terms, $1$ and $P_{B,m}^{j,(C,2)}\gamma_{SI}^j$, since they are negligible.
\begin{figure*}[!t]
\begin{equation}
\begin{aligned}
\label{e6}
\Gamma_{UC} = \frac{P_{m,B}^{j,(C,2)} P_{k,m}^{i,(C,1)} \alpha_{k,m}^i \beta_{m,B}^j} {1 + P_{m,B}^{j,(C,2)} \beta_{m,B}^j + P_{k,m}^{i,(C,1)}\alpha_{k,m}^i + P_{B,m}^{j,(C,2)}\gamma_{SI}^j + P_{B,m}^{j,(C,2)}P_{k,m}^{i,(C,1)}\gamma_{SI}^j\alpha_{k,m}^i} .\\
\end{aligned}
\end{equation}
\hrulefill
    \vspace*{4pt}
    \end{figure*}

Also, the SINR for non-cooperative communication is as follows\\
\begin{equation}
\begin{aligned}
\label{e7}
&\Gamma_{UNC1} = \frac{P_{m,B}^{i,(NC,1)} \beta_{m,B}^i} {1 + P_{B,m}^{i,(NC,1)} \gamma_{SI}^i}, &\\
&\Gamma_{UNC2} = \frac{P_{m,B}^{j,(NC,2)} \beta_{1,m,B}^j} {1 + P_{B,m}^{j,(NC,2)} \gamma_{SI}^j} , \\
\end{aligned}
\end{equation}

where, \\
\begin{equation}
\begin{aligned}
\label{e8}
&\alpha^i_{k,m} = \frac{|h^i_{k,m}|^2}{N_0W}, &\\
&\beta^j_{m,B} = \frac{|g^j_{m,B}|^2}{N_0W}, T = 1, &\\
&\beta^j_{1,m,B} = \frac{|g^j_{m,B}|^2}{N_0W}, T = 2, &\\
&\gamma^j_{SI} = \frac{|H^j_{SI}|^2}{N_0W},&
\end{aligned}
\end{equation}

where, $T$ represents the time slot.\\ 
Using the above SINRs, sum-rate of the network in cooperative and non-cooperative modes are derived as follows\\
\begin{equation}
\begin{aligned}
\label{e9}
%&R = \log_2(1+SINR)&\\
&1)R_{k,m}^{i,j,(C)} = \frac{1}{2} \log_2(1+\Gamma_{UC}),\\
&2)R_{m}^{i,(NC,l)} = \frac{1}{2} \log_2(1+\Gamma_{UNCl}), l=1,2.\\
%&R_{m}^{j,(NC,2)} = \frac{1}{2} \log_2(1+\Gamma_{UNC2}) .\\
\end{aligned}
\end{equation}

Total sum-rate is equal to ($\ref{e10}$) that is shown as follows\\
%\begin{figure*}[!t]
\begin{equation}
\begin{split}
\label{e10}
R_T &= \sum_{m=1}^{K2} \sum_{k=1}^{K1} \sum_{i=1}^N \sum_{j=1}^N \rho_{k,m}^{i,j} R_{k,m}^{i,j,(C)} \\
&\quad + \sum_{m=1}^{K2} \sum_{i=1}^N \sigma_{m}^{i,(1)} R_{m}^{i,(NC,1)} \\
&\quad + \sum_{m=1}^{K2} \sum_{j=1}^N \sigma_{m}^{j,(2)} R_{m}^{j,(NC,2)} .\\
\end{split}
\end{equation}
%\hrulefill
%    \vspace*{4pt}
%   \end{figure*}

In ($\ref{e10}$), $\rho$ and $\sigma$ are subcarrier indicators, that are equal to 0 or 1.
\section{Optimization Problem Formulation}
Based on what we have seen before; our optimization problem can be written as\\
\begin{equation}
\begin{aligned}
\label{e11}
%&maximize_{powers and subcarriers} R_T&\\
\displaystyle{\max_{P,\rho,\sigma} R_T}&\\
%&subject to:&\\
\textrm{s.t.}\\
&\displaystyle 1)\rho_{k,m}^{i,j},\sigma_{m}^{i,(1)},\sigma_{m}^{j,(2)} \in{(0,1)} \forall{k,m,i,j}, &\\
&2)\sum_{m=1}^{K2} \sum_{k=1}^{K1} \sum_{i=1}^N \rho_{k,m}^{i,j} + \sum_{m=1}^{K2} \sigma_{m}^{i,(1)} = 1 , \forall{i}, &\\
&3)\sum_{m=1}^{K2} \sum_{k=1}^{K1} \sum_{j=1}^N \rho_{k,m}^{i,j} + \sum_{m=1}^{K2} \sigma_{m}^{j,(2)} = 1 , \forall{j}, &\\
&4)P_{k,m}^{i,(C,1)}, P_{m,B}^{j,(C,2)}, P_{m,B}^{i,(NC,1)}, P_{m,B}^{j,(NC,2)}, P_{B,m}^{j,(C,2)}, &\\
&\quad P_{B,m}^{i,(NC,1)}, P_{B,m}^{j,(NC,2)} \geq 0 , \forall{k,m,i,j}, &\\
&5)\sum_{k=1}^{K1} \sum_{j=1}^N \rho_{k,m}^{i,j} (P_{k,m}^{i,(C,1)}+P_{m,B}^{j,(C,2)}) \leq Pmax^C_{k,m}, &\\
&6)\sum_{l=1}^N \sigma_{m}^{l,(1)} P_{m,B}^{l,(NC,1)} \leq Pmax^{NC}_{m}, l = i, j,  &\\
&7)\sum_{m=1}^{K2} \sum_{j=1}^N \sum_{i=1}^N \rho_{k,m}^{i,j} R_{k,m}^{i,j,(C)} \geq R_{k}^{(C),min}, &\\
&8)\sum_{l=1}^N \sigma_{m}^{l,(1)} R_{m}^{l,(NC,1)} \geq R_{m}^{(NC),min}, l = i, j .&\\
\end{aligned}
\end{equation}
%%%%%%%%%%%%%%%%hereeeeee
In the above problem, 1 to 3 are subcarrier constraints. $\sigma_{m}^{i,(1)}$ and $\sigma_{m}^{j,(2)}$ in the constraint 1 show that a subcarrier is assigned or not assigned to the user and $\rho_{k,m}^{i,j}$ shows that two subcarriers are assigned or not assigned to two users.
The second and third constraints represent that a user only can experience one type of communication in each time slot and at each time slot only one user can communicate. The constraints 4 to 6 are power constraints that show powers are positive and below a maximum specified power in cooperative and non-cooperative communications. The constraints 7 and 8 are data rate constraints. These constraints show that the data rate of each user is more than a minimum requirement in cooperative and non-cooperative communications. 

%\subsection{Solution Approach}
\subsection*{Solution Approach}
In this section, we prove that our optimization problem in ($\ref{e11}$) is a convex problem. First, we investigate the concavity of the uplink data rate of cooperative mode. Later we will discuss about the non-cooperative mode. To prove the concavity of data rate $R_c = \log(f)$, we need to show that the Hessian matrix of $f = 1+\Gamma_{UC}$ is negative semidefinite. That is, the eigenvalues of this Hessian matrix must be non-positive. Since the logarithm function is a concave and increasing function, if $f$ is concave, $R$ is concave too. Since the first term in (10) is concave (by properties of perspective operation), by the notations introduced in the table ($\ref{t1}$), we formulate the denominator of (6) as $a*c*x+b*y$. Then, the Hessian matrix of $f$ is as follows\\
\begin{equation}
\begin{aligned}
\label{e12}
& \begin{pmatrix}
  \frac{2a^3bc^2xy}{(by+acx)^3}-\frac{2a^2bcy}{\sigma_2} & \sigma_1 \\
  \sigma_1 & \frac{2ab^3xy}{(by+acx)^3}-\frac{2ab^2x}{\sigma_2} \\
 \end{pmatrix}, &\quad \\ 
& \text{where,} &\quad \\ 
&  \sigma_1 = \frac{ab}{by+acx}-\frac{ab^2y}{\sigma_2}-\frac{a^2bcx}{\sigma_2}+\frac{2a^2b^2cxy}{(by+acx)^3}, &\quad \\
 &\sigma_2 = (by+acx)^2 . &\\
\end{aligned}
\end{equation}

\begin{table}[t]
\caption{The notations that used in the Hessian matrix calculation}
\centering
\label{t1}
\begin{tabular}{|c|c|} 
\hline
 $P_{k,m} = x$ &$P_{m,B} = y$ \\
\hline
$\alpha_{k,m} = a$ &$\beta_{m,B} = b$ \\
\hline
$P_{B,m} \gamma_{SI} = c$ &$P_{B,m} = z$ \\
\hline
\end{tabular}
\end{table}

The determinant of ($\ref{e13}$) is zero, and its eigenvalues are as follows\\
\begin{equation}
\left\{ 0 , -\frac{2ca^2b^2x^2+2ca^2b^2y^2}{a^3c^3x^3+3a^2bc^2x^2y+3ab^2cxy^2+b^3y^3} \right\} .\\
\label{e13}
\end{equation}

The eigenvalues calculated above, are non-positive. It is proved that our function is concave and negative semidefinite. Now, we study the convexity of the uplink data rate equation of non-cooperative mode. We need to calculate the Hessian matrix of the data rate function in this mode to prove that it is a concave function. That is, \\
\begin{equation}
\Bigg( -\frac{b^2}{ln(2)(\frac{bx}{cz+1}+1)^2(cz+1)^2} \Bigg) .\\
\label{e14}
\end{equation}

The determinant of the above Hessian matrix is zero. Now, we calculate its eigenvalues\\
\begin{equation}
\left\{ 0,-\frac{b^2}{ln(2)(bx+cz+1)^2} \right\} .\\
\label{e15}
\end{equation}

It is shown that the eigenvalues are non-positive. It is proved that our function is concave and negative semidefinite. Now, we can use standard numerical solution methods for convex optimization problems to solve this problem.

\section{Numerical Results}
In this section, we represent results of the simulations that have been performed to evaluate the performance of the proposed schemes. We consider a single cell and a BS that is located at its center. We also have two groups of users that are uniformly distributed around the BS. The users of the first group are placed in a longer distance from the BS and the users of the second group in a shorter distance. All channel coefficients between users, between user and the BS, and also self-interference channels are independent and identically distributed complex Gaussian random variables with unit variance and zero mean. The users maximum power is assumed $P_{maxU} = 20 $dBm and the BS maximum power is $P_{maxBS} = 10 $dBW. We also assume that the bandwidth of each subcarrier is $20$kHz and thermal noise power is $-174 $dBm/Hz. The number of subcarriers is considered as default $N = 8$.

\begin{figure}[!h]
\begin{center}
\centering
\includegraphics[width=3.8in]{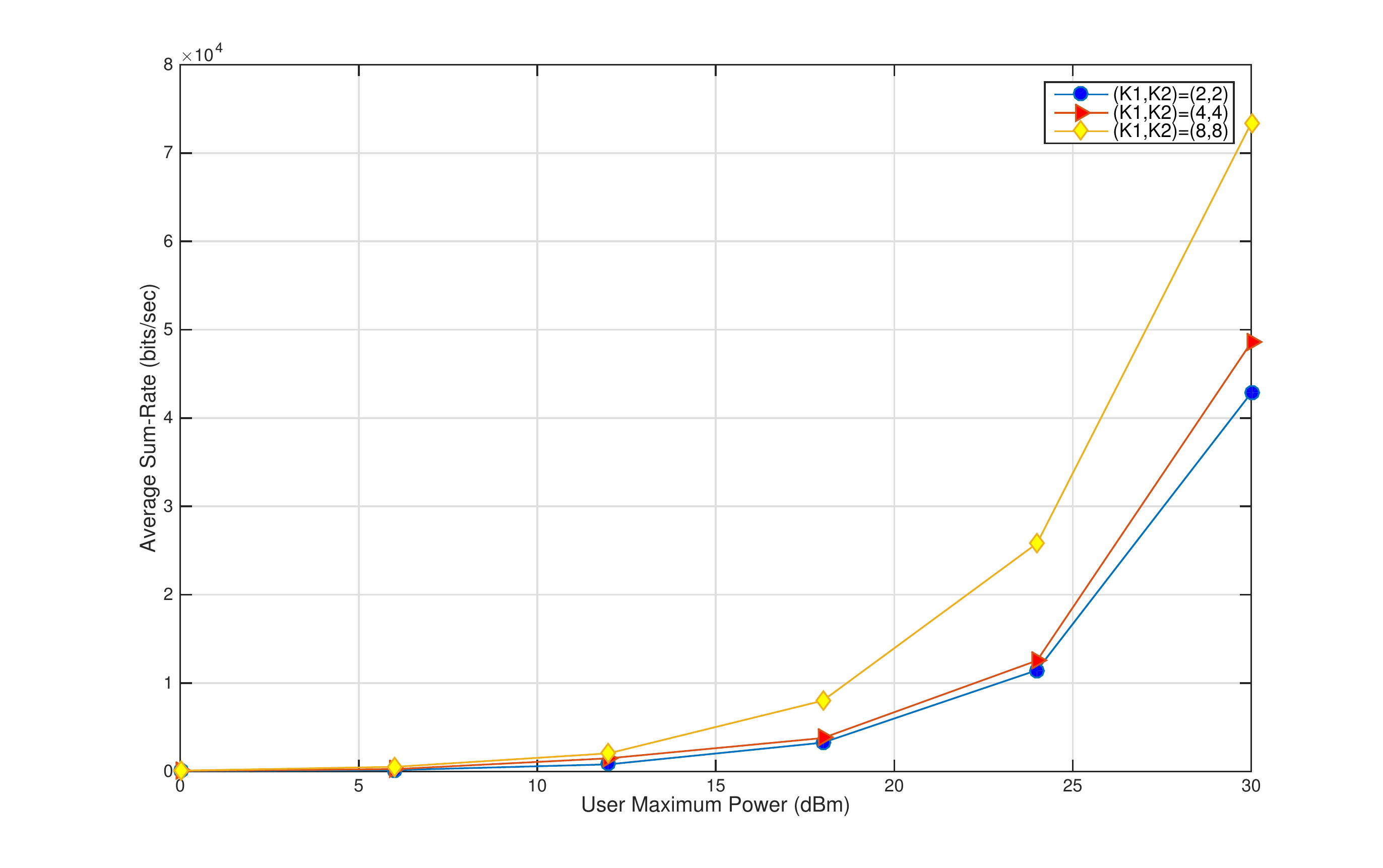}
\caption{Average sum-rate vs. user maximum power with self-interference.}
\label{f2}
\end{center}
\end{figure}

\begin{figure}[!h]
\begin{center}
\centering
\includegraphics[width=3.8in]{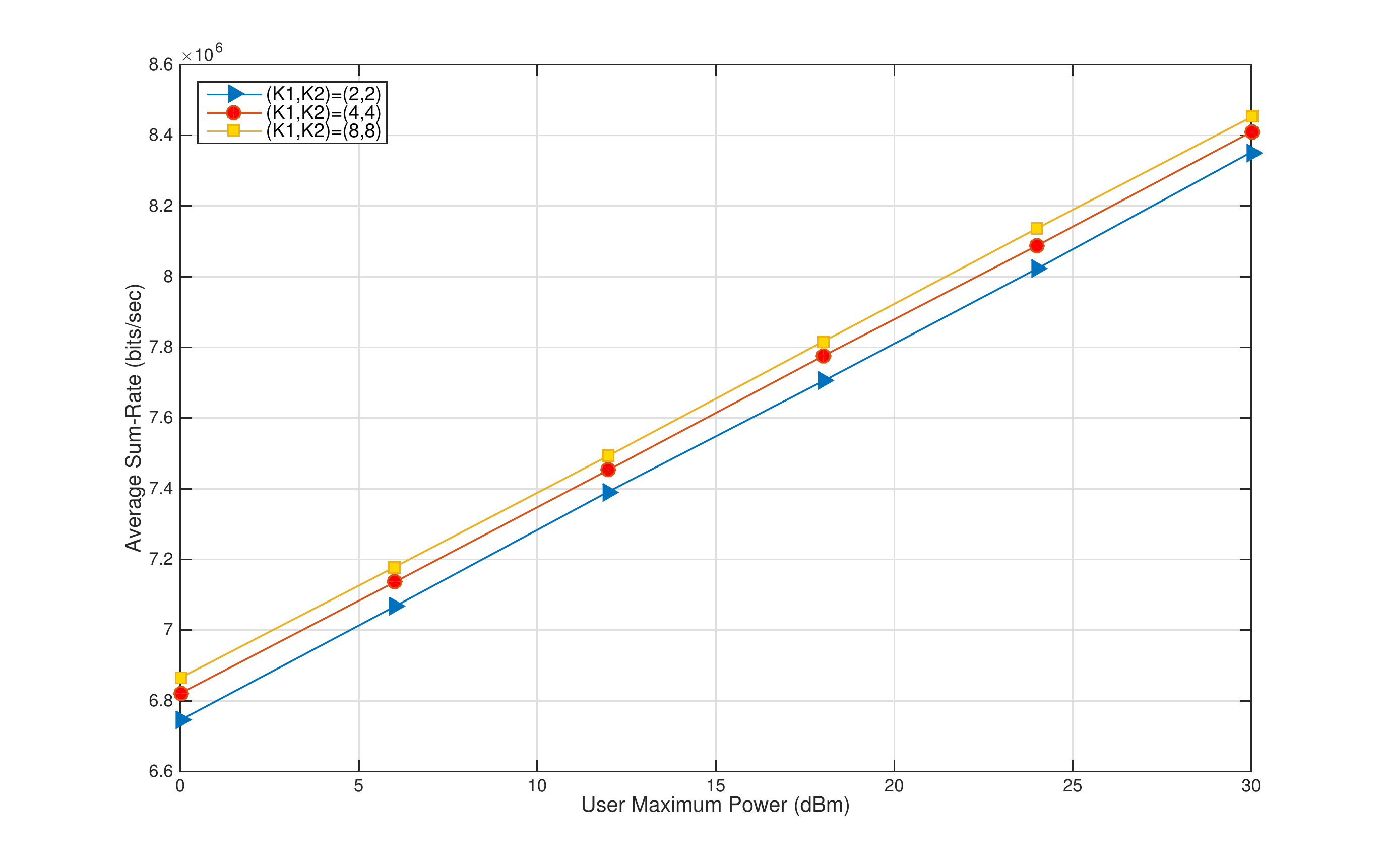}
\caption{Average sum-rate vs. user maximum power without self-interference.}
\label{f3}
\end{center}
\end{figure}

\begin{figure}[!h]
\begin{center}
\centering
\includegraphics[width=3.8in]{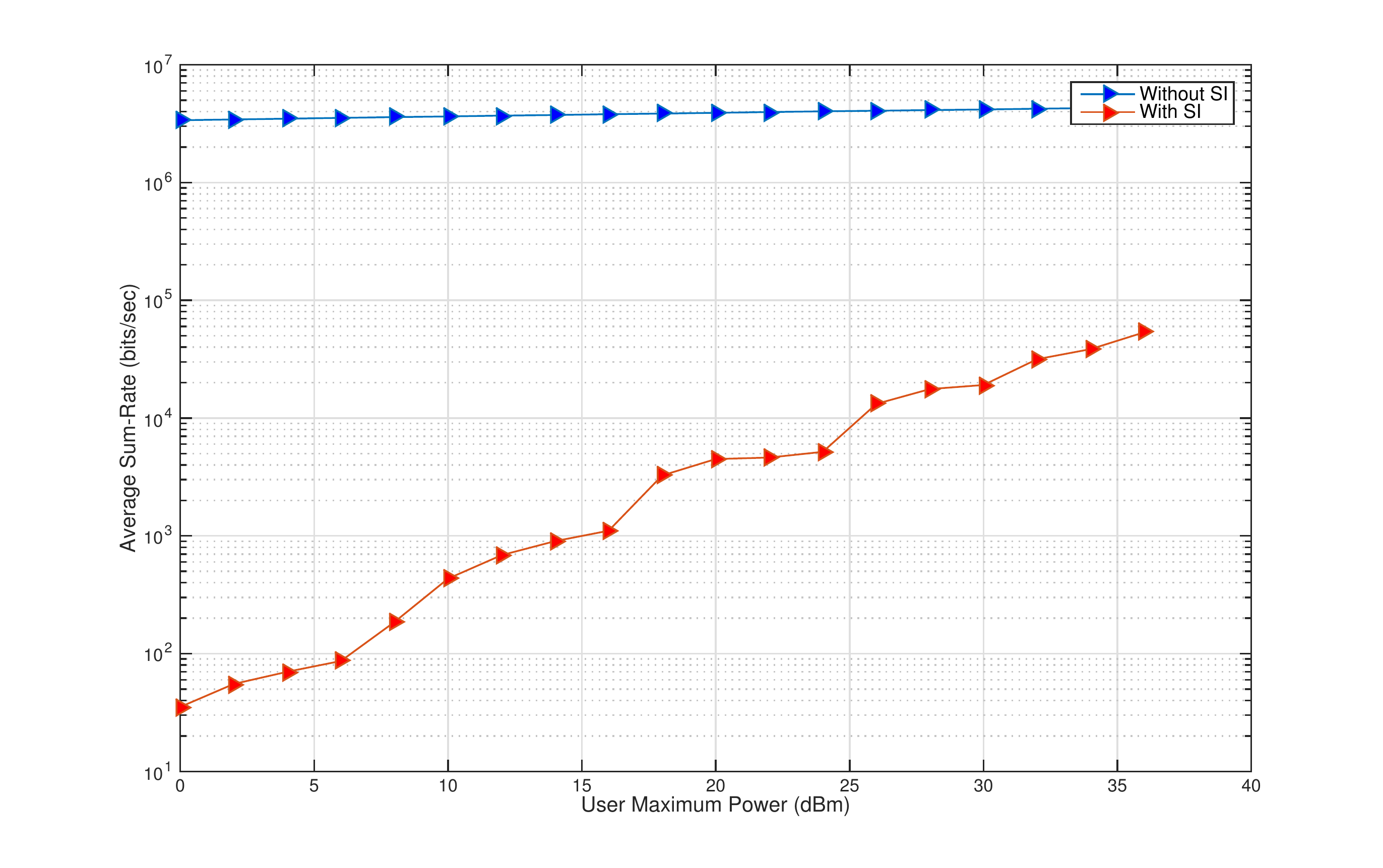}
\caption{Average sum-rate vs. users maximum power for comparison of self-interference effect.}
\label{f4}
\end{center}
\end{figure}

Fig. $\ref{f2}$ illustrates average sum-rate of users versus different values of user maximum power with consideration of self-interference effect for different K1 and K2 values. In this figure, our proposed scheme satisfies the QoS and power requirements. It is seen that as maximum power increases, the sum-rate increases and with increasing of K1 and K2, sum-rate increases too. Also, when we increase users maximum power, SINR can be more and then, sum-rate can be more. Fig. $\ref{f3}$, illustrates the average sum-rate of users versus different values of the user maximum power without self-interference effect for different K1 and K2 values. Like Fig. $\ref{f2}$,  with increasing the maximum power or the number of users, sum-rate increases. In Fig. $\ref{f4}$, average sum-rate versus users maximum power for two cases are presented. In the first case, the BS self-interference effect is considered, while in the second case, it is not considered. Our simulations illustrate that a network with more relays has more sum-rate in higher powers. In lower maximum power, the effect of users' power in calculation is very low and channel coefficient and noise and interference have more effect. By increasing the maximum power of users, the power effect in calculations can be more and we cannot ignore it. In general, we can say that equal number of users and the relays is the best.

\section{Conclusion}
In this paper, we considered the problem of sum-rate maximization in an IBFD Cooperative OFDMA Systems. We optimized power and subcarrier allocation by assuming maximum power and quality of service constraints. Then, we develop a new method to select the best relays among some users that these relays help other users with longer distance to the BS. The relays are selected with the criterion of maximum rate and using selective relaying. Also, we develop a new method for subcarrier allocation that is a Linear Assignment Problem Jonker-Volgenant Algorithm. In our work, we consider that the FD BS has self-interference that influence on data rate values. Our simulation results demonstrated that the proposed scheme performs better in terms of sum-rate compared to the existing system models in literature.

\end{document}